\documentclass[showpacs,twocolumn,preprintnumbers,
superscriptaddress,amsmath]{revtex4}

\usepackage{epsfig}

\newcommand{\PRL}[3]{Phys.\ Rev.\ Lett.\ {\bf #1},\ #2 (#3)}

\newcommand\g{\gamma}

\newcommand\m{\mu}

\newcommand{\be}{\begin{equation}}
\newcommand{\ee}{\end{equation}}
\newcommand{\bea}{\begin{eqnarray}}
\newcommand{\eea}{\end{eqnarray}}
\newcommand{\ba}[1]{\begin{array}{#1}}
\newcommand{\ea}{\end{array}}

\newcommand{\bm}[1]{\mbox{\boldmath${#1}$}}

\newcommand{\uk}{\hat{\mathbf{k}}}

\newcommand{\vg}{\bm{\gamma}}

\begin{document}

\title{Fate of the inert three-flavor, spin-zero color-superconducting phases}

\author{H.\ Malekzadeh}
\email{malekzadeh@figss.uni-frankfurt.de}
\affiliation{Frankfurt International Graduate School for Science,
J.W. Goethe-Universit\"at, D-60438 Frankfurt/Main, Germany}
\date{\today}

\begin{abstract}
I investigate some of the inert phases in three-flavor, spin-zero
color-superconducting quark matter: the CFL phase (the analogue of the
B phase in superfluid $^3\rm He$), the A and A* phases, and the 2SC and sSC phases.
I compute the pressure of these phases with and without the neutrality
condition. It is shown that the 2SC phase is identical
to the A* phase up to a color rotation. The CFL phase is the energetically
favored phase except for a small region of intermediate densities where the 2SC/A*
phase is favored.
\end{abstract}

\maketitle

\section{Introduction} \label{intro}

The interaction between electrons resulting from virtual exchange
of phonons is attractive when the energy difference between the
electrons states involved is less than the phonon energy. There
is a favorable attractive channel which dominates over the
repulsive screened Coulomb interactions and produces
superconductivity \cite{BCS}. In cold and dense quark matter, due
to asymptotic freedom \cite{asympt}, at quark chemical potentials
$\mu \gg \Lambda_{QCD}$ single-gluon exchange is the dominant
interaction between quarks and it is attractive in the
color-antitriplet channel. This leads to the formation of quark
Cooper pairs. This kind of condensation creates a new type of
superconductivity which is called color superconductivity (CSC).
The difference of this type of superconductivity compared to an
ordinary electronic conductor comes from the fact that quarks
carry different flavors and non-Abelian color charges \cite{CSC,2sc}.
Then, CSC can appear in different phases depending on the various
colors and flavors of the quarks which participate in Cooper
pairing.

It has been found that at asymptotically large baryon number
densities, where the masses of the $u$, $d$, and $s$ quarks are much
smaller than the chemical potential, the ground state of 3-flavor
QCD is the color-flavor-locked (CFL) phase \cite{CFL1}. In this
phase, quarks of all colors and flavors form Cooper pairs, and
$SU(3)_{c}\otimes SU(3)_L\otimes SU(3)_R\otimes U(1)_{B}$ is spontaneously 
broken to a subgroup $SU(3)_{c+f}\otimes U(1)_{c+em}$. This phase 
is a superfluid but an electromagnetic superconductor.
When the strange quark does not participate in pairing, 
the ground state is the so-called 2SC phase \cite{2sc}.
The original symmetry $SU(3)_{c}\otimes SU(2)_L\otimes SU(2)_R\otimes
U(1)_{em}\otimes U(1)_{B}$ breaks down to $SU(2)_{c}\otimes
SU(2)_L\otimes SU(2)_R\otimes U(1)_ {c+em}\otimes U(1)_{em+B}$. 
Unlike the CFL phase, the 2SC phase is not a superfluid.

In nature color superconductivity may exist in compact stars.
The approximation of asymptotically large density is not valid then. 
Moreover, the matter in the bulk of compact stars is neutral
with respect to color and electric charges. Besides, this
matter should remain in $\beta$-equilibrium \cite{neutrality1, 
neutral, gapless, phase-d}. These conditions impose stress on the system. 
In contrast to the 2SC and the CFL phases at large densities, in this case
 the Fermi momenta of different quark flavors participating in pairing 
 cannot be equal. Therefore, the ground state is neither a pure CFL nor a
pure 2SC state. There might appear phases in which the Cooper
pairs carry nonzero total momenta and the system exhibits a
crystalline structure due to a spatially varying energy gap. This
kind of a superconductor was for the first time studied by Larkin
and Ovchinnikov \cite{LO} and Fulde and Ferrell \cite{FF}
which is called LOFF superconductor. Alternately,
there might be ungapped quasiparticle excitations with
nonzero gap parameter for both 2SC and CFL phases, giving rise to so-called
gapless phases \cite{neutral, gapless}, but it has been shown that gapless
phases are unstable \cite{gaplessinstabilities}. In Ref.\cite{kaon}, the
authors argued that the system may respond to this stress by forming a
kaon condensate in the ground state. With this condensation the
strange quark number density is decreased without a costly breaking
of the pairs in the CFL background. 

From the theoretical point of view, superfluidity is very similar to 
superconductivity. Both are states of interacting many-fermion systems
that are distinguished from the normal matter by an order parameter. 
Among all possible phases of superfluid $^3\rm He$ states only the 
so-called inert phases have been found in experiment. These phases
play a crucial role in superconductivity due to Michel's theorem,
cf.\;Ref.\cite{michel, inert}. It is known that the B phase
is the dominant phase in a vast area of the phase diagram of the
superfluid $^3\rm He$, while the so-called A phase has a dominant
pressure only in a tiny area of the diagram. 

The CSC for one flavor is rather similar to superfluid $^3\rm He$.
The reason is that the order parameters are quite similar.
In one-flavor CSC, the order parameter is a $3 \times 3$
matrix in color and spin space. For comparison, in superfluid
$^3\rm He$, the order parameter is a $3 \times 3$ matrix in 
coordinate and spin space. It is natural to expect that 
one-flavor CSC has inert phases which are rather similar to 
the ones in superfluid $^3\rm He$. These inert phases have
already been studied in Ref.\cite{spinn0}.

In three-flavor, spin-zero CSC
the order parameter is a $3 \times 3$ matrix in color 
and flavor space. In this case, it was shown in 
Ref.\ \cite{andkrish} that there are 511 possible phases 
but using the symmetries of the system this number could be 
lowered. 
Naturally, some of these are analogues of the inert phases 
encountered in one-flavor CSC and superfluid $^3\rm He$. 
The CFL phase is the analogue of the CSL phase in one-flavor 
CSC or the B phase in superfluid $^3\rm{He}$. The 2SC phase is the 
analogue of the polar phase in one-flavor CSC or superfluid 
$^3\rm He$. There is, however, also an analogue of the A 
phase of one-flavor CSC and superfluid $^3\rm{He}$, and a 
closely related phase, the A* phase which I discuss here 
for the first time. Finally, another inert phase is the sSC 
phase which is the complement of the 2SC phase in the sense 
that strange quarks pair with up and down quarks, but up 
and down quarks do not pair among themselves.

As it was mentioned earlier, at asymptotically large 
densities, it is known \cite{CFL1} that the CFL phase is 
the energetically favored phase. At intermediate densities, 
the 2SC phase may be energetically preferred over the
CFL phase, depending on the value of the quark masses or 
the diquark coupling strength \cite{2sc}. The purpose of 
this paper is to answer the question whether some of the 
other inert phases could be energetically favored over 
either the CFL or the 2SC phase at intermediate densities.

The paper is organised as follows. In the next section, I 
give a general expression for the pressure
of the different phases without imposing the neutrality
condition. Then, I compare the pressure of all phases to 
determine the ground state. In Sec.{III}, Iđ calculate 
the pressure of the phases including the neutrality condition. 
In order to identify the chemical potentials that make the 
system color and charge neutral, I evaluate the tadpoles
in each phase. At the end, I consider the symmetry breaking 
pattern and give a summary.

\section{Pressure without the neutrality condition}\label{pwn}

In this section I calculate the pressure of CSC quark matter at 
zero temperature without the neutrality condition. In the absence 
of the neutrality condition such an analysis is easy in the 
framework of QCD. This is because there is a single chemical 
potential for all quarks. 

To calculate the difference
between the pressure of a color-superconducting phase and the
normal conducting phase, I follow the general approach of
 \;Refs.\cite{spinn0, spin0}. Without going into details, here 
 I present only the main points of the formalism.

For condensation in the even-parity channel and in the
ultrarelativistic limit I use the ansatz introduced for the gap
matrix in Eq.\,(2.24) of Ref.\cite{spinn0},
\be\label{gmatrix}
\Phi^+(K)\,=\,\sum_{e=\pm}\,\phi_e(K)\,{\cal M}^-
\Lambda^e_{\bf k}\,\,,
\ee
where $\phi_e(K)$ is the gap function, $\Lambda_{\bf k}^e\equiv
(1+e\g_0\vg\cdot\uk)/2$ are projectors onto
the positive $({\rm e}=+1)$ and negative $({\rm e}=-1)$ energy
states, and ${\cal M}^-$ is defined
by the order parameter $\Delta^k_h$,
\be\label{mdelta}
{\cal M}^-\,=\,I_k\Delta^k_h J^h\gamma_5\; ,
\ee
with the following expressions,
\be\label{levicivita}
I_k\equiv(I_{ij})_k \hspace{.5cm},\hspace{.5cm}J^h\equiv(J^{fg})^h
\ee
which are completely antisymmetric $3 \times 3$ matrices, constructed
from the Levi-Civita tensor $-i\,\epsilon_{ijk}$. In spin-one CSC $I_k$
is an antisymmetric matrix in color space and $J^h$ is an antisymmetric
matrix in spin space. The $k$ and $h$ indices are for the associated
color and spin components respectively.
${\cal M}^-$ satisfies
\be \label{M}
[{\cal M}^-,\Lambda_{\bf k}^e]=0 \,\, .
\ee
To proceed I introduce new quantities $L^+$ in terms
of ${\cal M}^-$ and the Dirac matrix $\gamma_{0}$ via,
\be \label{lmatrix}
L^+\,\equiv\,\gamma_0\,{\cal M}^+ \,{\cal M}^-\,
\gamma_{0} \,\, ,
\ee
where ${\cal M}^+ = \gamma_0 ({\cal M}^-)^\dagger \gamma_0$. The matrix
$L^{+}$ can be expanded in terms of projectors ${\cal P}^{r}$
with eigenvalues $\lambda_{r}$,
\be\label{pmatrix}
L^{+}\,=\,\lambda_{r}\,{\cal P}^{r}\,\,.
\ee
This imposes the following form on the projectors ${\cal P}^{r}$,
\be\label{plmatrix}
{\cal P}^{r}\,=\,\prod_{s\neq r}^n\frac{L^+ - \lambda_s}
{\lambda_r - \lambda_s}\,\,,
\ee
where $n$ is the number of different eigenvalues.
Using the ansatz in Eq.\,(\ref{gmatrix}) in the QCD gap equation, 
one finds the following zero-temperature result for the value of 
the gap $\phi_0$ at the Fermi surface \cite{spinn0}
\be\label{phi0}
\phi_0 = 2\,\, b \, b_0' \, \m \, \exp\left(- \frac{\pi}{2 \,\bar{g}}
\right)\, \left( \langle\lambda_1\rangle^{a_1} \,
\langle\lambda_2\rangle^{a_2}\, \langle\lambda_3\rangle^{a_3}
\right)^{-1/2}\,.
\ee
In this equation, $a_1$, $a_2$, and $a_3$ are positive constants defined by
\be
a_s\,=\,\frac{n_s\lambda_{s}}{\sum_r\,n_r\lambda_{r}}\,\, ,
\ee
where $n_s$ is the degeneracy of eigenvalue $\lambda_s$. They obey
the constraint
\be\label{condition}
\sum_{s} a_s\,=\,1\,\,.
\ee
The remaining constants in Eq.\,(\ref{phi0}) are as follows,
\be \label{constants}
\tilde{b} \equiv 256 \pi^4\left(\frac{2}{N_f g^2} \right)^{5/2}\,\, ,
\,\, b_0' \equiv\exp\left(-\frac{\pi^2+4}{8}\right) \,\, ,
\ee
where $N_f$ accounts for the number of flavors that occur,
$g$ is the strong coupling constant, and
\be \label{b}
b\equiv \tilde{b}\,\exp(-d)\,\,.
\ee
The constant $d$ originates from subleading contributions
to the gap equation. For spin-zero condensates, due to an accidental
cancellation of some of the subleading terms arising from static
electric and non-static magnetic gluon exchange, $d$ is zero. In the
spin-one cases, this cancellation does not occur and $d\neq0$,
cf.\;Ref.\,\cite{d-nonzero}.

Finally, at zero temperature and without color and electric
charge neutrality conditions, the difference between the pressure of
the color-superconducting phase and the normal conducting phase is given
by \cite{spinn0}
\be\label{pressure1}
\Delta\,p\,=\,\frac{\mu^2}{16\pi^2}\,\phi^2_0\,
{\rm Tr}[L^+]\,\,.
\ee

In three-flavor, spin-zero color superconductivity the order parameter
$\Delta^k_h$, cf.\, Eq.\,(\ref{mdelta}), is a $3 \times 3$ matrix in
color $(k = 1, 2, 3)$ and flavor $(h = 1, 2, 3)$ space. 
Different order parameters $\Delta^k_h$ lead
to different gap matrices $\Phi^+$, and thus to 
different physical states. As we see from Eq.\,(\ref{pressure1})
the value of the pressure is given in terms of ${\rm Tr}[L^+]$ and the value 
of the gap. Then, different order parameters also produce in general
different values for the pressure. In the following, after introducing the
order parameter of each phase I calculate their pressure.

\subsection{A phase}

The order parameter of the so-called A phase has the following
form, cf.\,Refs.\cite{spinn0,inert},
\be\label{A-phase}
\Delta^k_h\,=\,\delta^{k3}(\delta_{h1}+i\delta_{h2})\,\,.
\ee
where the upper index stands for color and the lower
index for flavor. Using this equation the matrix ${\cal M}^-$ becomes
\be\label{m-a}
{\cal M}^-\,=\,I_3(J^1 + iJ^2)\gamma_5\,\,.
\ee
Inserting this expression in Eq.\,(\ref{lmatrix}) leads to
\bea
[L^+]^{fg}_{ij} &=& (\delta_{ij}-\delta_{i3}\,\delta_{j3})(2\,
\delta^{fg}-\delta^{f1}\,\delta^{g1}\nonumber\\
&-&\delta^{f2}\,\delta^{g2}-i
\delta^{f2}\,\delta^{g1}+i\delta^{f1}\,\delta^{g2})\,\,.
\eea
One should notice that because of
the summation rule in Eq.\,(\ref{mdelta}), the role of the indices
is interchanged, so that the upper index stands for flavor and the lower
index for color. After some straightforward calculation, one finds that,
\be\label{root}
[L^{+}]^n\,=\,2^{n-1}\,[L^{+}]\,\,.
\ee
The results for the eigenvalues $\lambda_{r}$ come from the
roots of the following equation,
\be
{\rm det}(\lambda-L^{+})\,=\,0\,\, ,
\ee
The left-hand side of this equation can be rewritten in the form
\be
{\rm det}(\lambda-L^{+})\,=\,\rm{exp}\,\{\rm{Tr}\,[\rm{ln}
(\lambda-L^{+})]\}\,\,.
\ee
which, after expanding the logarithm and making use of 
Eq.\,(\ref{root}), gives
\be
\lambda^5(\lambda-2)
^4\,=\,0\,\,.
\ee
This equation yields two different eigenvalues for the A phase,
\be\label{eigen-a}
\bigg\{\begin{array}{ccc} \lambda_1\,=\,2&\hspace{.4cm}
(4-{\rm fold})&\rightarrow a_1=1\;,\\ \lambda_2\,=
\,0&\hspace{.4cm}(5-{\rm fold})&
\rightarrow a_2=0\,\,. \end{array}
\ee
From Eqs.\,(\ref{phi0}) and (\ref{pressure1}) I find the
value of the pressure for this phase,
\bea\label{pressureA}
\Delta\,p_{\rm A}\,=\,4\,\alpha,
\eea
where $\alpha$ is defined as
\be
\alpha\,=\,\frac{\mu^4}{4\pi^2}\, b^2 \, b_0'\,^2 \, \exp
\left(- \frac{\pi}{\bar{g}}
\right)\,.
\ee

\subsection{A* phase}

Here I define a new phase motivated by the order parameter
of the A phase and I call it A* phase. This phase
is not included in Ref.\cite{spinn0} but was introduced in Ref.\cite{inert}.
The order parameter of this phase is a transposed form of that
in the A phase, i.e., the roles of the color and flavor indices
are interchanged,
\be\label{A*-phase}
\Delta^k_h\,=\,(\delta^{k1}+i\delta^{k2})\delta_{h3}\,\,.
\ee
By the same argument which led to this phase one
realizes that the corresponding $L^+$ matrix can be
derived by interchanging the color and flavor indices
of the matrix $L^+$ in the A phase,
\bea
[L^+]^{fg}_{ij}&=&(2\,\delta_{ij}-\delta_{i1}\,\delta_{j1}-
\delta_{i2}\,\delta_{j2}-i\delta_{i2}\,\delta_{j1}\nonumber\\
&+&i\delta_{i1}\,
\delta_{j2})(\delta^{fg}-\delta^{f3}\,\delta^{g3})\,\,.
\eea
Therefore, one has the same expression for $[L^+]^n$ and the same
eigenvalues $\lambda_{1,2}$ given for the A phase, cf.\,Eqs.\,
(\ref{root}) and (\ref{eigen-a}) respectively. Considering
Eqs.\,(\ref{phi0}) and (\ref{pressure1}), the pressure of
this phase is given by
\be
\Delta\,p_{{\rm A^{\star}}}\,=\,4\,\alpha\,\,.
\ee
which is equal to the pressure of the A phase.

\subsection{Planar or sSC phase}\label{planar}

Another experimentally observed phase in superfluid $^3\rm{He}$ is 
the so-called planar phase which has the following form for the order 
parameter
\be
\Delta^k_h\,=\,\delta^k_h-\delta^{k3}\delta_{h3}\,\,,
\ee
where $\Delta^3_3=0$. This form of the order parameter 
corresponds to the so-called sSC phase in three-flavor 
CSC, cf.\,Ref.\,\cite{pd-nu}, 
where the pressure of that phase was calculated including the neutrality 
condition. After some straightforward but tedious
calculation the $L^+$ matrix for this phase
is found to be
\bea
[L^+]^{fg}_{ij}&=&2\,\delta_{ij}\delta^{fg}-
\delta^{fg}(\delta_{i1}\delta_{j1}+\delta_{i2}\delta_{j2})
\nonumber\\
&-&\delta_{ij}(\delta^{f1}\delta^{g1}+\delta^{f2}\delta^{g2})
\nonumber\\
&+&(\delta_{i1}\delta^{f1}-\delta_{i2}\delta^{f2})
(\delta_{j1}\delta^{g1}-\delta_{j2}\delta^{g2})\,\, ,
\eea
which gives the following result
\be
[L^{+}]^n\,=\,2^{n-1}\,L^{+}+(2^{n-1}-1)\ell
\,\, ,
\ee
where $\ell$ is
\bea
[\ell]^{fg}_{ij}&=& 2\,(\delta_{i1}\delta^{f1}
-\delta_{i2}\delta^{f2})(\delta_{j1}\delta^{g1}-\delta_{j2}
\delta^{g2})
\nonumber\\
&-&\delta^{fg}(\delta_{i1}\delta_{j1}+\delta_{i2}\delta_{j2})
\nonumber\\
&-&\delta_{ij}(\delta^{f1}\delta^{g1}+\delta^{f2}\delta^{g2})\,.
\eea
Then following the method introduced for calculating the eigenvalues 
of the A phase I derive
\be\label{eigen}
\bigg\{\begin{array}{ccc} \lambda_1\,=\,2&\hspace{.4cm}
(2-{\rm fold})&\rightarrow a_1=1/2\, ,\\
\lambda_2\,=\,1&\hspace{.4cm}(4-{\rm fold})&
\rightarrow a_2=1/2\, ,\\
\lambda_3\,=\,0&\hspace{.2cm}\hspace{.3cm}(3-{\rm fold})&
\rightarrow\hspace{.2cm} a_3=0\,\,. \end{array}
\ee
The difference between the pressure of the sSC phase and
normal conducting matter is found to be
\bea
\Delta\,p_{{\rm sSC}}\,=\,\frac{8}{2^{1/2}}\,\alpha\,\,.
\eea

\subsection{Polar or 2SC phase}

Analogous to the previous Sec. \ref{planar}, the phase called
the polar phase for superfluid $^3\rm He$ is analogous to
the 2SC phase in CSC,
\be
\Delta^k_h\,=\,\delta^{k3}\delta_{h3}\,\, ,
\ee
which gives a zero value for the gaps
$\Delta^1_1$ and $\Delta^2_2$. The $L^+$ matrix of
this phase is
\be
[L^+]^{fg}_{ij}\,=\,(\delta_{ij}-\delta_{i3}\,\delta_{j3})
(\delta^{fg}-\delta^{f3}\,\delta^{g3})\,\,,
\ee
and yields
\be
[L^+]^n\,=\,L^+\,\, ,
\ee
so that with the same methods one arrives at
\be\label{eigen}
\bigg\{\begin{array}{ccc} \lambda_1\,=\,1&\hspace{.4cm}
(4-{\rm fold})&\rightarrow a_1=1\;,\\
\lambda_2\,=\,0&\hspace{.4cm}(5-{\rm fold})&
\rightarrow a_2=0\;, \end{array}
\ee
with the pressure difference equal to
\bea
\Delta\,p_{{\rm 2SC}}\,=\,4\,\alpha\,\,.
\eea

\subsection{CFL phase}
To make a detailed comparison with the previous results I copy
the results given for the CFL phase from Ref.\,\cite{spinn0}.
The order parameter of the CFL phase is
\be
\Delta^k_h\,=\,\delta^k_h\,\, ,
\ee
and the $L^+$ matrix has the following form,
\be
[L^+]^{fg}_{ij}\,=\,\delta^f_i\,\delta^g_j+\delta^{fg}\,\delta_{ij}\;,
\ee
with the following quantities,
\be\label{eigen-cfl}
\bigg\{\begin{array}{ccc} \lambda_1\,=\,4&\hspace{.4cm}
(1-{\rm fold})&\rightarrow a_1=1/3\,\,,\\ \lambda_2\,=\,1&\hspace{.4cm}
(8-{\rm fold})&\rightarrow a_2=2/3\,\,, \end{array}
\ee
which are sufficient to find the pressure of this phase,
\bea
\Delta\,p_{\,{\rm CFL}}\,=\,\frac{12}{2^{1/3}}\,\alpha\;.
\eea

Using all results, one can compare the pressure of the inert phases
\be\label{resulttt}
P_{{\rm CFL}}\,>\,P_{{\rm sSC}}\,>\,P_{{\rm A}}\,=\,P_{{\rm A^*}}\,=
\,P_{{\rm 2SC}}\,\,.
\ee

As we see, the pressure of the CFL phase is larger than the
pressure of the other phases, i.e., the CFL phase is the 
dominant phase. Another
interesting result is a larger value for the pressure
of the sSC phase without the neutrality
condition than that for the 2SC phase.

In the next section I calculate the pressure including
the neutrality condition. As it was mentioned
above, in this case the pressure for some of these
inert phases (2SC, sSC, and CFL) were considered in the
literature, cf.\,Ref.\,\cite{already1,blaschke}.
Therefore, in the following I calculate the
pressure of only those phases which are still left out,
the A and A* phases.

\section{Pressure including the neutrality condition}
\label{pwnc}

In this section I impose the neutrality condition on the system. 
For this one has to know the relevant chemical potentials of each phase. 
Thus one has several parameters to be evaluated and using QCD 
to calculate the pressure becomes more complicated. 
Therefore, I will use the Nambu-Jona-Lasinio model (NJL), cf.\cite{NJL}.

To find the pressure of the A and the A* phase under the
neutrality condition using the NJL model one has to know the 
relevant chemical potentials for these phases. Since a 
nonvanishing tadpole leads to the violation of the neutrality 
of the system, one has to introduce a chemical potential 
which makes the tadpole vanish \cite{tadpoles,tad}. The sum of 
these chemical potentials together with the quark chemical 
potential is the relevant chemical potential for the system 
under color and electric charge neutrality condition. 
Thus, before I proceed to evaluate the pressure I compute tadpoles 
of the system. Afterwards, I go further to find the pressure of 
the systems.

\subsection{Calculating the tadpoles}

In order to compute the tadpoles of a system I use Eq.\,(19)
of Ref.\,\cite{tadpoles},
\be\label{tadpole}
  \mathcal{T}^a=-{g\over2}\int {d^4Q\over i(2\pi)^4}\mathrm{Tr}_
  {D,c,f}[\Gamma_0^a
  G^+(Q)+\bar\Gamma_0^aG^-(Q)]\,\,.
\ee
Here $\Gamma_0^a=\gamma_0 T^a$, $\bar\Gamma_0^a=-\gamma_
0 (T^a)^T$ with  $T^a= \,\lambda^a/2$ for $a=1,\ldots,8$
where $\lambda^a$ are the Gell-Mann matrices in flavor space,
and $G^\pm$ are the fermion propagators for
quasiparticles and charge-conjugate quasiparticles,
\be\label{sigma}
G^{\pm}(Q)\,\equiv\,([G^{\pm}_0]^{-1}-\Sigma^{\pm})^{-1}\hspace{.2cm},
\hspace{.2cm}\Sigma^{\pm}\,\equiv\,{\cal M}^{\mp}\,G^{\mp}_0\,{\cal M}^{\pm}\;.
\ee
In these equations, $\Sigma^{\pm}$ is the quark self-energy generated
by exchanging particles or charge conjugate particles with the
condensate. The role of different phases appears in the quark self-energy
via the matrix ${\cal M}^{\mp}$, cf.\,Eq.(\ref{mdelta}). In the next subsections,
\ref{tad-a} and \ref{tad-a*}, I insert the order parameter of each phase
into ${\cal M}^{\mp}$ to find the tadpoles.

\subsection{Tadpoles in the A phase}\label{tad-a}

In Eq.(\ref{tadpole}) I first evaluate the
trace over color and flavor space, and afterwards the trace over the
Dirac space. The inverse free fermion propagator for quarks $([ G_0^\pm ]^{-1})^
{fg}_{ij}$ has the following color and flavor structure,
\be\label{propag}
([ G_0^\pm ]^{-1})^{fg}_{ij} = (\gamma^\mu K_\mu
\pm \hat \mu \gamma_0 - \hat{M})\delta^{fg}\delta_{ij} \; ,
\ee
Using Eq.\,(\ref{m-a}) yields
\bea
(\Sigma^{\pm})^{fg}_{ij}&\equiv&({\cal M}^{\mp}\,G^{\mp}_0\,{\cal M}^{\pm})
^{fg}_{ij}=\gamma_{5}\,\frac{\Delta^2}{\,\,\gamma^\mu k_\mu \mp {\hat\mu}
\gamma_0-{\hat{\cal M}}\,\,} \nonumber\\
&\times&
\gamma_5 \,(\delta^{fg}-
\delta^{f 3}\delta^{g 3})(2\,\delta_{ij}-\delta_{i1}\delta_{j1}
-\delta_{i2}\delta_{j2} \nonumber\\
&\pm& 
i\delta_{i2}\delta_{j1} \mp i\delta_{i1}\delta_{j2})\,\, ,
\eea
where $\Delta$ is the value of the gap for this phase. After some algebraic
calculation, one can find the color and flavor structure of the propagator
$G^{\pm}(Q)$ as,
\bea
[G^{\pm}]^{fg}_{ij}\,&=&\,([G_{0}^{\pm}]^{-1}-2\gamma_5\,G^{\mp}
\gamma_5\Delta^2)^{-1}(\delta^{fg}-\delta^{f3}\delta^{g3})
\delta_{ij}\nonumber\\
&+&
[G_0^{\pm}]\delta^{f3}\delta^{g3}\delta_{ij}-([G_{0}^{\pm}]^{-1}
-2\gamma_5\,G^{\mp}\gamma_5\Delta^2)^{-1}\nonumber\\
&\times&
(\gamma_5\,G^{\mp}\gamma_5\Delta^2)[G_{0}^{\pm}](\delta^{fg}-\delta^{f3}
\delta^{g3})\nonumber\\&\times&(\delta_{i1}\delta_{j1}+\delta_{i2}
\delta_{j2}\pm i\delta_{i1}\delta_{j2} \mp i\delta_{i2}\delta_{j1})\,\, .
\eea
Now one has to put these equations back into Eq.\,(\ref{tadpole}) and evaluate
the traces. By doing so, it is revealed that all components of
$\mathcal{T}^a$ are zero except for $a=8$. This forces us to introduce a
chemical potential $\mu_8$ for the system to make the tadpole vanish and
achieve color and electric charge neutrality.

\subsection{Tadpoles in the A* phase}\label{tad-a*}

Following the same procedure for calculating the tadpoles of the A phase,
the color and flavor structure of the propagator $G^{\pm}(Q)$ is given by
\bea
[G^{\pm}]^{fg}_{ij}\,&=&\,([G_{0}^{\pm}]^{-1}-2\gamma_5\,G^{\mp}
\gamma_5\Delta^2)^{-1}(\delta_{ij}-\delta_{i3}\delta_{j3})\delta^{fg}
\nonumber\\
&+&[G_0^{\pm}]\delta_{i3}\delta_{j3}\delta^{fg}
-([G_{0}^{\pm}]^{-1}-2\gamma_5\,G^{\mp}\gamma_5\Delta^2)^{-1}
\nonumber\\
&\times&
(\gamma_5\,G^{\mp}\gamma_5\Delta^2)[G_{0}^{\pm}](\delta_{ij}
-\delta_{i3}\delta_{j3})\nonumber\\
&\times&(\delta^{f1}\delta^{g1}+\delta^{f2}
\delta^{g2}\pm i\delta^{f1}\delta^{g2} \mp i\delta^{f2}
\delta^{g1})\,\, .
\eea
This form of the full fermion propagator yields a nonzero value
for the $a=2$ and $a=8$ tadpoles.
Hence, for this phase the chemical potential has to contain $\mu_2$ and
$\mu_8$ to provide neutrality.

Now one is able to calculate the pressure.

\subsection{Pressure}

The grand partition function is given by
\be\label{Z}
\mathcal{Z} \equiv {\rm e}^{-\Omega V/T}
= \int \mathcal{D} \bar\psi \mathcal{D} \psi \, {\mathrm e}^{i
\int_X \left( \mathcal{L} + \bar\psi \hat{\mu} \gamma^0 \psi
\right) } \; ,
\ee
where $\Omega$ is the thermodynamic potential density, $V$ is the
volume, $\hat\mu$ is the matrix of the quark
chemical potentials and $\mathcal{L}$ is the Lagrangian density
for three-flavor quark matter which for a local NJL-type interaction
is given by
\bea\label{Lagrangian}
\mathcal{L} &=& \bar \psi \, ( i \partial\hspace{-.2cm}/ - \hat{m} \, )
\psi +G_S \sum_{a=0}^8 \left[ \left( \bar \psi \lambda_a \psi \right)^2
+ \left( \bar \psi i \gamma_5 \lambda_a \psi \right)^2 \right]
\nonumber \\
&+& G_D \sum_{k,h} \left[\bar{\psi}_{i}^{f} i \gamma_5
\epsilon^{ijk}
\epsilon_{fgh} (\psi_C)_{j}^{g} \right] \left[
(\bar{\psi}_C)_{\rho}^{r} i \gamma_5
\epsilon^{\rho \sigma k} \epsilon_{rsh} \psi_{\sigma}^{s}
\right]
\nonumber \\
&-& K \left\{ \det_{F}\left[ \bar \psi \left( 1 + \gamma_5 \right) \psi
\right] + \det_{F}\left[ \bar \psi \left( 1 - \gamma_5 \right) \psi
\right] \right\} \;,
\eea
where the quark spinor field $\psi_{i}^{f}$ carries color
($i=r,g,b$) and flavor ($f=u,d,s$) indices. The matrix of quark
current masses is given by $\hat{m} = {\rm diag}_{F}(m_u, m_d, m_s)$ and
$\lambda_0\equiv \sqrt{2/3} \,\openone_{F}$. The charge-conjugate
spinors are defined as $\psi_C = C \bar \psi^T$ and $\bar
\psi_C = \psi^T C$, where $\bar\psi=\psi^\dagger \gamma^0$
is the Dirac-conjugate spinor and $C=i\gamma^2 \gamma^0$
is the charge conjugation matrix. Note that I include the 't Hooft
interaction whose strength is determined by the coupling constant
$K$. This term breaks the $U(1)$ axial symmetry.

The term in the second line of Eq.\,(\ref{Lagrangian}) describes a
scalar diquark interaction in the color-antitriplet and 
flavor-antitriplet channel. For symmetry reasons there should also be
a pseudoscalar diquark interaction with the same coupling constant
but for the sake of simplicity I do not consider it here.

I use the following set of model parameters \cite{RKH}:
\bea\label{model-parameters}
m_{u,d} &=& 5.5 \; \mathrm{MeV} \; , \\
m_s &=& 140.7 \; \mathrm{MeV} \; , \\
G_S \Lambda^2 &=& 1.835 \; , \\
K \Lambda^5 &=& 12.36 \; , \\
\Lambda &=& 602.3 \; \mathrm{MeV} \; .
\label{Lambda}
\eea
In general, it is expected that the diquark coupling $G_D$ is
of the same order as the quark-antiquark coupling $G_S$ and in this
paper, I study the regime of intermediate coupling strength with
 $G_D=\frac34 G_S$.

All quarks carry baryon charge $1/3$ and thus have a diagonal contribution 
$\mu \delta^{fg}\delta_{ij}$ to their matrix of chemical potentials. By 
definition $\mu = \mu_{B}/3$ with $\mu_{B}$ being the baryon chemical
potential. In order to fulfil the neutrality condition one has to 
take the chemical potentials introduced in
the previous sections, cf.\,\ref{tad-a} and \ref{tad-a*}, into account. 
For the A phase one has
\be\label{chem-a}
\mu_{ij}^{fg} = \left(
  \mu \delta^{fg}
+ \mu_Q Q_{F}^{fg} \right)\delta_{ij}
+ \mu_8\left(T_8\right)_{ij}\delta^{fg} \; ,
\ee
and for the A* phase
\be\label{chem-a*}
\mu_{ij}^{fg} = \left(
  \mu \delta^{fg}
+ \mu_Q Q_{F}^{fg} \right)\delta_{ij}
+ \left(\mu_2\left(T_2\right)_{ij} + \mu_8
\left(T_8\right)_{ij}\right)\delta^{fg} \; ,
\ee
where $\mu_Q$ is the chemical potential of the electric
charge and $Q_{F}$ is the electric charge matrix 
$Q_{F}=\mbox{diag}_{F}(\frac23,-\frac13,-\frac13)$.

To calculate the mean-field thermodynamic potential
at temperature $T$, one has to linearize the interaction in the
presence of the diquark condensates $\Delta^k_h \sim
(\bar{\psi}_C)_{i}^{f} i \gamma_5 \epsilon^{ijk}
\epsilon_{fgh} \psi_{j}^{g}$ and the quark-antiquark
condensates $\sigma_\alpha \sim \bar \psi_\alpha^a
\psi_\alpha^a$ (no sum over $\alpha$). Then, integrating out the
quark fields and neglecting the fluctuations of composite order
parameters gives the following expression for the
thermodynamic potential:
\bea\label{Omega}
\Omega &=& \Omega_{L} +\frac{1}{4 G_D} \sum_{k,h=1}^{3}
\left| \Delta^{k}_{h}\right|^2 +2 G_S \sum_{\alpha=1}^{3}
\sigma_\alpha^2 \nonumber\\
&-& 4 K \sigma_u \sigma_d \sigma_s
-\frac{T}{2V} \sum_K \ln \det \frac{S^{-1}}{T} \; ,
\eea
Here I added the lepton contribution $\Omega_{L}$. The inverse
full quark propagator $S^{-1}$ in the Nambu-Gorkov representation is
\be
S^{-1} =
\left(
\begin{array}{cc}
[ G_0^+ ]^{-1} & \cal{M}^- \\
\cal{M}^+ & [ G_0^- ]^{-1}
\end{array}
\right) \;.
\label{off-d}
\ee
The constituent-quark mass matrix in the inverse propagator
of quarks and charge-conjugate quarks, cf.\,Eq.(\ref{propag}),
is defined as
$\hat{M} = {\rm diag}_{F}(M_u,M_d,M_s)$ with
\be
M_\alpha = m_\alpha - 4 G_S \sigma_\alpha
+ 2 K \sigma_\beta \sigma_\gamma \; ,
\label{Mi}
\ee
where the set of indices $(\alpha, \beta, \gamma)$ is a permutation of
$(u,d,s)$.

The off-diagonal components of the inverse propagator in
Eq.\,(\ref{off-d}) are given in Eq.\,(\ref{mdelta}) in terms of
the diquark condensates $\Delta^{k}_{h}$. The color- and flavor-symmetric 
condensates are neglected here because they are small
and not crucial for the qualitative understanding of the phase
diagram \cite{phase-d}.

The inverse quark propagator is a $72\times 72$ matrix. It was
shown in the Appendix A of Ref.\,\cite{already1} that for a
real-valued order parameter this matrix
contains a two-fold spin and Nambu-Gorkov degeneracy. Then, 
it is sufficient to evaluate its
$18$ nondegenerate eigenvalues $\epsilon_i$,
\be\label{s-1}
   \det \frac{S^{-1}}{T}
=  \prod_{i=1}^{18}
   \left( \frac{ \omega_n^2 + \epsilon_i ^2 }{T^2}
   \right)^2 \; .
\ee
However, for a complex-valued order parameter, which is the
case for the A and A* phases, cf.\,Eqs.\,(\ref{A-phase})
and (\ref{A*-phase}), one has to decompose the inverse propagator
matrix into a real and an imaginary part to use the simplest
numerical recipes. This doubles the dimension of the complex
propagator matrix. Since one has to find the determinant of this
matrix one can transform the matrix to a block-diagonal form
without losing anything. After that, one separates the nondegenerate
eigenvalues and follow the same procedure for
a real-valued order parameter. I find the
same degeneracies for the inverse quark propagator as in 
the case of real-valued order parameters. 
Therefore at the end, the number of the non-degenerate
eigenvalues $\epsilon_i$ decreases again to $18$. With
$p\equiv -\Omega$ I find
\bea
p &=& \frac{1}{2 \pi^2} \sum_{i=1}^{18} \int_0^\Lambda {\mathrm d} k \, k^2
\left[ |\epsilon_i| + 2 T \ln \left( 1 + {\mathrm e}^{-
\frac{|\epsilon_i|}{T}} \right) \right] \nonumber \\
&+& 4 K \sigma_u \sigma_d \sigma_s
- \frac{1}{4 G_D} \sum_{k,h=1}^{3} \left| \Delta^k_h \right|^2
-2 G_S \sum_{\alpha=1}^{3} \sigma_\alpha^2
\nonumber \\
&+& \frac{T}{\pi^2} \sum_{l=e,\mu} \sum_{\epsilon=\pm}
\int_0^\infty {\mathrm d} k \, k^2
\ln \left( 1 + {\mathrm e}^{-\frac{E_l-\epsilon\mu_l}{T}}\right)\; ,
\label{pressure}
\eea
where the contribution of electrons and muons with masses $m_e \approx
0.511$ MeV and $m_\mu \approx 105.66$ MeV are included.
The expression for the pressure in Eq.\,(\ref{pressure}) has a physical
meaning only when the chiral and color-superconducting order parameters,
$\sigma_\alpha$ and $\Delta^k_h$, satisfy the following set of equations:
\bea\label{gapeqns}
\frac{\partial p}{\partial \sigma_\alpha} &=& 0\;,\\
\frac{\partial p}{\partial \Delta^k_h} &=& 0\;.
\eea
In order to enforce the condition of local charge neutrality
in dense matter, for the A phase one has to require that
\bea
n_Q &\equiv& \frac{ \partial p }{\partial \mu_Q} = 0\;,\\
n_8 &\equiv& \frac{ \partial p }{\partial \mu_8} = 0\;,
\eea
and for the A* phase
\bea
n_Q &\equiv& \frac{ \partial p }{\partial \mu_Q} = 0\;,\\
n_2 &\equiv& \frac{ \partial p }{\partial \mu_2} = 0\;,\\
n_8 &\equiv& \frac{ \partial p }{\partial \mu_8} = 0\;.
\eea
By these equations the chemical potentials
$\mu_Q$, $\mu_2$, and $\mu_8$ are fixed, but the quark
chemical potential $\mu$ is left as a free parameter.
\begin{figure}
\begin{center}
\includegraphics[width=7cm]{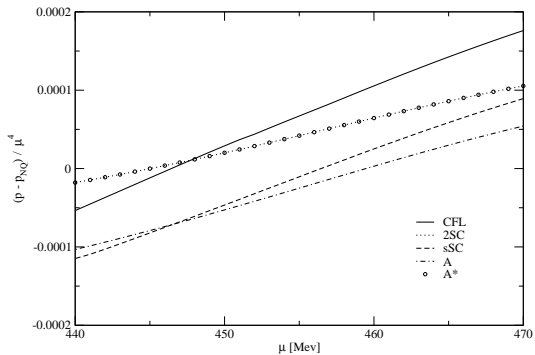} \vspace{.1cm}
\caption{Pressure divided by $\mu^4$ for the different neutral
solutions of the gap equation at $T\,=\,0$ as a function of the
quark chemical potential $\mu$. The diquark coupling strength is
$G_D\,=\,\frac{3} {4}\,G_S$.} \label{plot}
\end{center}
\end{figure}

The difference between the pressure $p$ of different phases and 
the pressure of normal quark matter $p_{NQ}$ versus the chemical 
potential $\mu$ at zero temperature is depicted in Fig.\,\ref{plot}. 
One observes that at large chemical potentials the
pressure for the A and the A* phases is less than that for the
CFL phase. Also the pressure of the sSC phase, which in the case
that I did not enforce the neutrality condition was larger
than that for the 2SC phase, is smaller in this case. Therefore,
in this circumstances the CFL phase is again the ground state
of matter. Furthermore, the pressure of the A phase is smaller than 
that for the other phases. At smaller chemical potentials the 2SC/A* 
phase wins over the CFL phase.

Another interesting result comes from comparing the pressure
of the 2SC and the A* phase. One observes that as before,
cf.\,Eq.\,(\ref{resulttt}), their pressures are exactly
equal for all values of the chemical potential. The question
arises whether the A* phase is a different form of the 2SC
phase? To answer that, one has to consider the order parameter
of these phases and find a unitary transformation between
these two phases. The order parameter in three-flavor,
spin-zero color superconductivity is a $3\times 3$ matrix
in color and flavor space. Because of the different masses of
the quark flavors one cannot find a unitary matrix in this space
which transforms the A*(2SC) phase matrix into the 2SC (A*)
phase matrix, but it is possible to investigate this in color
space. The spinor fields in the Lagrangian,
Eq.\,(\ref{Lagrangian}), are changed under a color
transformation $U={\rm exp}(-i\omega^\alpha\lambda^\alpha)$ as,
\bea
\psi\,&\rightarrow&\,U\,\psi\,\, , \\
\overline{\psi}\,&\rightarrow&\,\overline{\psi}\,U^\dagger\,\, , \\
\psi^c\,&\rightarrow&\,U^*\,\psi^c\,\, ,\\
\overline{\psi^c}\,&\rightarrow&\,\overline{\psi^c}\,U^T\,\, ,
\eea
with which the expression in the second line of
Eq.\,(\ref{Lagrangian}),
\bea
\sum_{k,h}(\bar{\psi}_C)_{\rho}^{r} i \gamma_5
\epsilon^{\rho \sigma k} \epsilon_{rsh} \psi_{\sigma}^{s}&\equiv&
(\bar{\psi}_C)_{\rho}^{r} i \gamma_5 (\epsilon^{\rho\sigma 1}
+i\epsilon^{\rho\sigma 2})\epsilon_{r s\,3} \psi_{\sigma}^{s}\nonumber\\
&=&
(\bar{\psi}_C)_{\rho}^{r} i \gamma_5(\lambda_5+i\lambda_7)^
{\rho\sigma}\epsilon_{r s\,3} \psi_{\sigma}^{s}\; ,\nonumber\\
\eea
transforms into
\bea
(\bar{\psi}_C)_{\rho}^{r} i \gamma_5\,U^T\,(\lambda_5+i\lambda_7)
^{\rho\sigma}U\epsilon_{r s\, 3} \psi_{\sigma}^{s}\;.
\eea
Hence, one has to find a unitary matrix $U$ which changes the last
expression to that in the 2SC phase,
\be
(\bar{\psi}_C)_{\rho}^{r} i \gamma_5\,(\lambda_2)^{\rho\sigma}\,
\epsilon_{r s\,3} \psi_{\sigma}^{s}\,\, .
\ee
The relevant matrix is
\be\label{transform}
U={\rm e}^{i\pi(\lambda_5+\lambda_6)/2\sqrt{2}}\,\, ,
\ee
by which one has
\be
U^T\,(\lambda_5+i\lambda_7)U\,\rightarrow\,i\sqrt{2}
\lambda_2\,\, .
\ee
Using Eq.\,(\ref{Z}) one observes that
\be
\bar\psi \,U^\dagger\hat{\mu}^{A^*}\,U \gamma^0 \psi\,\rightarrow\,
\bar\psi \hat{\mu}^{2SC} \,\gamma^0 \psi\,\, ,
\ee
and the following results are derived,
\bea
\mu_2^{A^*}\,&\equiv&\,-\,\frac{\sqrt{3}}{2}\,\mu_8^{2SC}\,\, ,
\nonumber\\
\mu_8^{A^*}\,&\equiv&\,-\,\frac{1}{2}\,\mu_8^{2SC}\,\, .
\eea

I conclude that the A* phase is the same as the 2SC phase.
This was not the case for superfluid $^3\rm He$,
cf.\,Ref.\,\cite{inert}.
 
In the next section I briefly mention the
result derived for the symmetry pattern of the A* and 2SC
phases and then investigate the generators of the symmetry groups
for the A* phase in terms of those in the 2SC phase.

\section{Pattern of symmetry breaking} \label{sym}

The study of the order parameter of the A* phase reveals that 
just like the 2SC phase the strange quark does not participate 
in pairing. Thus, the A* phase is a two-flavor CSC.
Without including the neutrality condition and in the case
that all the quark masses are zero, the initial symmetry
group for the A* phase is the same as the 2SC phase
\bea
G&=&SU(3)_{c}\otimes SU(2)_{L}\otimes SU(2)_R 
\nonumber\\&\otimes& U(1)_{B}\otimes U(1)_{em}\;,
\eea
where $SU(3)_{c}$ is the color gauge group and
$SU(2)_L$ and $SU(2)_R$ are the representations of the flavor
group. $U(1)_{B}$ and $U(1)_{em}$, respectively, are accounting for
baryon number conservation symmetry and the electromagnetic gauge group.
The order parameter $\Delta$ is an element of a representation of
$G$. After pairing, the group $G$ is spontaneously broken to a
residual subgroup $H\,\subseteq\,G$ so that any transformation
$g\,\in\,H$ leaves the order parameter invariant,
\be
g(\Delta)\,=\,\Delta\,\, .
\ee

To find all possible order parameters and the corresponding
residual groups $H$ one has to satisfy this invariance condition.
Here I restrict the calculations to those which lead to the
residual group of the A* phase. Using the method given in
Ref.\,\cite{inert, spin0} I find that
\bea\label{residual}
H_{A^*}&=&SU(2)_c\otimes SU(2)_L\otimes SU(2)_R
\nonumber\\&\otimes& \tilde{U}(1)_B\otimes\tilde{U}(1)_{em}\;,
\eea
which is exactly the same residual group as for the 2SC phase.
This result confirms the equivalence of the A* phase with the
2SC phase from this point of view. To complete this subsection, 
utilizing the results of
the previous section and knowing the generators for the residual
group of the 2SC phase, I want to find the corresponding
generators for the A* phase.
In the 2SC phase the generator of baryon number conservation is
\bea
\tilde{B}\,=\,B\,-\,\frac{2}{\sqrt{3}}T_8\,\, ,
\eea
and that for an unbroken $\tilde{U}(1)_{em}$ is
\be
\tilde{Q}\,=\,Q\,-\,\frac{1}{\sqrt{3}}T_8\,\, .
\ee
Under the same color transformation for which the A* phase goes
to the 2SC phase, Eq.\,(\ref{transform}), one finds the generators
for the residual group of the A* phase,
\bea
\tilde{B}'\,=\,B\,+(T_2\,+\,\frac{1}{\sqrt{3}}T_8)\,\, , \\
\tilde{Q}'\,=\,Q\,+\,\frac{1}{2}(T_2\,+\,\frac{1}{\sqrt{3}}T_8)\;,
\eea
which is a linear combination of generators of the 2SC phase.

\section{Conclusions}
I investigated the inert phases in three-flavor CSC, 
the A, A*, 2SC, sSC, CFL phases. Without the
neutrality condition, calculating the pressure of the phases
I found the CFL phase to be the dominant phase. Besides, in this case 
I found that the pressure of the sSC phase is larger than that
for the 2SC phase. Including the neutrality condition, the CFL
is again the dominant phase and the dominance of the sSC phase
over the 2SC phase ceases. On the other hand, I found that 
for all values of the chemical potential the pressure of
the A* and the 2SC phases are equal. This led me to show 
that the A* phase is different from the 2SC phase only by a 
color rotation. At the end I found the generators of the 
A* phase in terms of those for the 2SC phase.

Although none of the newly investigated phases are favored over the 
2SC or CFL phases, this does not preclude that they could not exist 
if the external conditions are changed. For instance, at zero 
temperature the A phase in superfluid $^3\rm He$ appears only 
if a sufficient external magnetic field is applied. 
The same also could happen in neutron stars which have strong 
magnetic fields. This could be 
an interesting subject for further studies.

\section{Acknowledgements}
The author thanks D.\ H. Rischke and I.\ Shovkovy for
very fruitful and prolific discussions, S.\ B.\ R\"uster
for providing the numerical codes. He also thanks the Frankfurt
International Graduate School for Science for support.

\end{document}